%% file: root.tex
\documentclass[letterpaper, 10 pt, conference]{ieeeconf}  

\IEEEoverridecommandlockouts                              

\usepackage{graphics} 
\usepackage{graphicx}
\usepackage{epsfig} 
\usepackage{mathptmx} 
\usepackage{times} 
\usepackage{amsmath} 
\usepackage[T1]{fontenc}
\usepackage{amsfonts,amssymb}
\usepackage{bm}
\usepackage {subcaption}

\usepackage{CJK}
\usepackage{algorithm}
\usepackage{algorithmicx}
\usepackage{algpseudocode}
\newcommand{\RNum}[1]{\uppercase\expandafter{\romannumeral #1\relax}}

\title{\LARGE \bf
Safe Multi-Agent Reinforcement Learning through Decentralized Multiple Control Barrier Functions
}

\author{Zhiyuan Cai$^{1,*}$, Huanhui Cao$^{1,*}$, Wenjie Lu$^{1}$, Lin Zhang$^{2}$, and Hao Xiong$^{1,\dag}$

\thanks{*Authors have contributed equally.}

\thanks{$^{1}$Zhiyuan Cai, Huanhui Cao, Wenjie Lu, and Hao Xiong are with the School of Mechanical Engineering and Automation, Harbin Institute of Technology Shenzhen, Shenzhen, China.
        }%
\thanks{$^{2}$Lin Zhang is with the Department of Aerospace Engineering and Engineering Mechanics, University of Cincinnati, Cincinnati, OH 45040, USA.
        }%
        
\thanks{$\dag$Corresponding author: e-mail: xionghao@hit.edu.cn.}
}

\begin{document}

\maketitle
\thispagestyle{empty}
\pagestyle{empty}

\begin{abstract}
Multi-Agent Reinforcement Learning (MARL) algorithms show amazing performance in simulation in recent years, but placing MARL in real-world applications may suffer safety problems. MARL with centralized shields was proposed and verified in safety games recently. However, centralized shielding approaches can be infeasible in several real-world multi-agent applications that involve non-cooperative agents or communication delay. Thus, we propose to combine MARL with decentralized Control Barrier Function (CBF) shields based on available local information. We establish a safe MARL framework with decentralized multiple CBFs and develop Multi-Agent Deep Deterministic Policy Gradient (MADDPG) to Multi-Agent Deep Deterministic Policy Gradient with decentralized multiple Control Barrier Functions (MADDPG-CBF). Based on a collision-avoidance problem that includes not only cooperative agents but obstacles, we demonstrate the construction of multiple CBFs with safety guarantees in theory. Experiments are conducted and experiment results verify that the proposed safe MARL framework can guarantee the safety of agents included in MARL.

\end{abstract}

\section{Introduction}

Reinforcement learning (RL) is passing from one step of success to another in solving challenging problems from robot control \cite{Hwangbo2019LearningRobots} to playing games \cite{Silver2016} in recent years. A number of real-world problems that can be addressed by RL, such as self-driving vehicles, drone swarms, as well as cooperation and competition of robots, are inherently multi-agent. Thus, multi-agent RL (MARL) approaches have been developed to deal with the growing joint action space and address decentralization constraints as a result of the spiking of agents \cite{Samvelyan2019TheChallenge}. MARL shows promising performance in simulation, but placing agents in real-world safety-critical applications can lead to challenges in safety for MARL \cite{Fisac2019ASystems,Cheng2020SafeUncertainties}.

Collision-avoidance is a fundamental challenge for MARL in real-world applications. For example, if the MARL is applied to self-driven vehicles, the failure in collision-avoidance can lead to injury or death of humans or damage of property \cite{Xiong2021SafetyControl}. However, relying on the reward functions of MARL only is not sufficient to guarantee safety \cite{ElsayedAly2021SafeMR}. The collision-avoidance problem has been addressed based on different approaches, such as velocity obstacles \cite{Wilkie2009GeneralizedObstacles}, Voronoi partitions \cite{Hu2020FormationPartition}, differential games \cite{Goode2012ALimitations}. Nonetheless, the computational expense and the conservativeness of these approaches limit their applications in MARL.

Safe RL takes the safety issue into account, aiming to learn a policy that maximizes the expected reward on the condition that safety constraints are satisfied \cite{Garcia2012SafeLearning}. Several safe RL approaches have been proposed, including learning from demonstration, policy optimization with constraints, and reward-shaping. However, the major issue of these approaches is that safety is not guaranteed during initial learning interactions \cite{Cheng2019End-to-endTasks}. To address this issue, shielding frameworks have been proposed, including shield RL \cite{Alshiekh2018SafeRL} and centralized shield MARL \cite{ElsayedAly2021SafeMR}. According to the shielding frameworks, shields are synthesized to guarantee safety during learning by monitoring the actions of agents. The concepts of shield RL and centralized shield MARL are decent. However, it can be challenging to design shields in real-world applications. For the centralized shield MARL, its scalability is limited by the centralized shield, and communication delay can limit its application in safety-critical problems in the real world.

Control Barrier Function (CBF) based methods can lead to effective shields in safety-critical control problems \cite{Ames2019ControlApplications}. Safe RL approaches based on integrating RL with shields based on CBFs have been proposed in \cite{Cheng2019End-to-endTasks,Marvi2020SafeApproach} as well. To ensure safety in multi-agent problems, multi-agent CBF has been developed in \cite{Borrmann2015ControlBehavior}. Different CBFs have been designed assuming that agents are cooperative, neutral, and competitive, respectively. Cheng et al. \cite{Cheng2020SafeUncertainties} have further improved the multi-agent CBF from the perspective of uncertainty bounds. However, CBFs have not been applied to MARL to guarantee the safety of MARL yet.

This paper integrates CBFs to MARL and achieves MARL with decentralized shields for the first time. The major contributions of this paper are as follows.

\begin{itemize}
\item This paper develops a safe MARL framework integrating decentralized cooperative CBFs and non-cooperative CBFs to agents of MARL in view of the fact that agents and entities in a multi-agent environment can be cooperative and non-cooperative. Based on the safe MARL framework, Multi-Agent Deep Deterministic Policy Gradient (MADDPG) \cite{Lowe2017Multi-agentEnvironments} is developed to Multi-Agent Deep Deterministic Policy Gradient with Control Barrier Functions (MADDPG-CBF). 
\item This paper shows that the proposed safe MARL framework is with safety guarantees, in theory, in certain applications. Through introducing the design of a cooperative CBF and a non-cooperative CBF in a collision-avoidance problem, it is shown that the proposed safe MARL framework can provide safety guarantees for an agent of MARL in theory.
\end{itemize}

The rest of the paper is organized as follows. Section II introduces the preliminaries of this paper, introducing MARL and CBF. In Section III, a safe MARL framework with decentralized multiple CBFs is proposed and MADDPG-CBF is developed from the MADDPG according to the safe MARL framework. The design of CBFs for an agent suffering cooperative agents and non-cooperative agents is demonstrated based on a collision-avoidance problem in Section IV. In Section V, the safety of the MADDPG-CBF is validated in a collision-avoidance problem. Finally, Section VI summarizes this paper.

\section{Preliminaries}

\subsection{Multi-Agent Reinforcement Learning}
MARL scales RL to environments with multiple agents. A MARL problem is often presented as a Markov game $(N,X,{\{U^i\}}_{i \in N},P,{\{r^i\}}_{i \in N},\gamma)$ with a finite set $N=\{1,2,...,n_A\}$ of agents \cite{ElsayedAly2021SafeMR}. $X$ represents the state space of the Markov game. $U^i$ represents the action space of agent $i \in N$. $P$ is the transition function of the Markov game. $r^i$ denotes an immediate reward function for agent $i$. $\gamma$ is a discount factor. At time step $t$, each agent chooses an action based on its observation. Then, the Markov game evolves from $x_t \in X$ to $x_{t+1} \in X$ and agent $i$ achieves a reward $r^i_t$. An individual agent $i$ aims to learn a policy that optimizes the cumulative rewards $\sum_{t=0}^{\infty}{\gamma^t r_t^i}$. 

MADDPG is a MARL algorithm proposed for environments with mixed cooperative and competitive. The MADDPG adopts an actor-critic framework of centralized training and distributed execution.
An agent has a critic-network that can access global information and an actor-network that can access local observation information only. The MADDPG has been applied to several problems (e.g., pursuit-evasion games \cite{Wang2020CooperativeLearning} and combat tasks \cite{Zhang2019EfficientTasks}). 


\subsection{Control Barrier Function}
A control barrier function plays a role in the study of safety equivalent to a Lyapunov function in the study of stability \cite{Ames2019ControlApplications}. Without loss of generality, one can assume a nonlinear affine system
\begin{equation}
\label{controlSystem}
\dot{x} = f(x) + g(x)u
\end{equation}with $f$ and $g$ locally Lipschitz, $x \in \mathbb{R}^n $ and $u \in U \subset \mathbb{R}^m$. Safety of the system can be guaranteed via enforcing the invariance of a safe set \cite{Ames2019ControlApplications}. In particular, we consider a set $C$ defined as the superlevel of a continuously differentiable function $h(x):\mathbb{R}^n \rightarrow \mathbb{R}$, yielding
\begin{equation}
\begin{split}
\label{setC}
C=\{x\in \mathbb{R}^n : h(x)\ge 0\}\\
\partial C=\{x\in  \mathbb{R}^n : h(x) = 0\}\\
Int(C)=\{x\in \mathbb{R}^n : h(x) > 0\}
\end{split}
\end{equation}
We refer to $C$ as the \emph{\underline{safe set}}.

\noindent \textbf{Definition 1.} \cite{Ames2019ControlApplications} \emph{The set $C$ is \underline{forward invariant} if for every $x(0) \in C, x(t) \in C $ for all $t \in [0, t_{max})$. The system (\ref{controlSystem}) is \underline{safe} with respect to the set $C$ if the set $C$ is forward invariant.}

 A CBF $B(x):C\rightarrow \mathbb{R}$ satisfying
\begin{equation}
    \inf_{x\in Int(C)}B(x)\ge 0, \qquad \lim_{x\rightarrow\partial C}B(x)=\infty
\end{equation}
was proposed in \cite{Prajna2007ACertificates}  and the conditions on $B(x)$ were relaxed by \cite{Ames2014ControlControl} as
\begin{equation}
\label{B}
    \dot{B}(x) \le \frac{\Gamma}{B(x)}
\end{equation}
where $\Gamma > 0$.

\noindent \textbf{Definition 2.} \cite{Ames2014ControlControl} \emph{For the system (\ref{controlSystem}), a function $B(x):C\rightarrow \mathbb{R}$ is a \underline{control barrier function} for the set $C$ defined by (\ref{setC}) for a continuously differentiable function $h(x):\mathbb{R}^n\rightarrow\mathbb{R}$, if there exist locally Lipschitz class $\kappa$ function $\alpha_1,\alpha_2$ such that, for all $x\in Int(C)$},
\begin{equation}
    \frac{1}{\alpha_1(h(x))} \le B(x) \le \frac{1}{\alpha_2(h(x))}
\end{equation}
\begin{equation}
    \inf_{u \in U}[L_fB(x)+L_gB(x)u-\frac{\Gamma}{B(x)}] \le 0
    \label{Bx}
\end{equation}
Given a CBF $B(x)$ and a set of Lipschitz continuous controller
\begin{equation}
    K_B(x) = \{u\in U:L_fB(x)+L_gB(x)u -\frac{\Gamma}{B(x)} \le 0 \}
\end{equation}
one has

\noindent \textbf{Theorem 1.} \cite{Ames2014ControlControl} \emph{Given a set $C\subset \mathbb{R}^n$ defined by (\ref{setC}) with associated control barrier function B(x), any Lipschitz continuous controller $u \in K_B(x)$ for the system (\ref{controlSystem}) renders the set $C$ \underline{forward invariant}, ensuring that the system (\ref{controlSystem}) is  \underline{safe} with respect to the set $C$.}



\section{Safe Multi-Agent Reinforcement Learning with Decentralized Control Barrier Functions}

\begin{figure}[t]
     \centering
     \includegraphics[width=8.5cm]{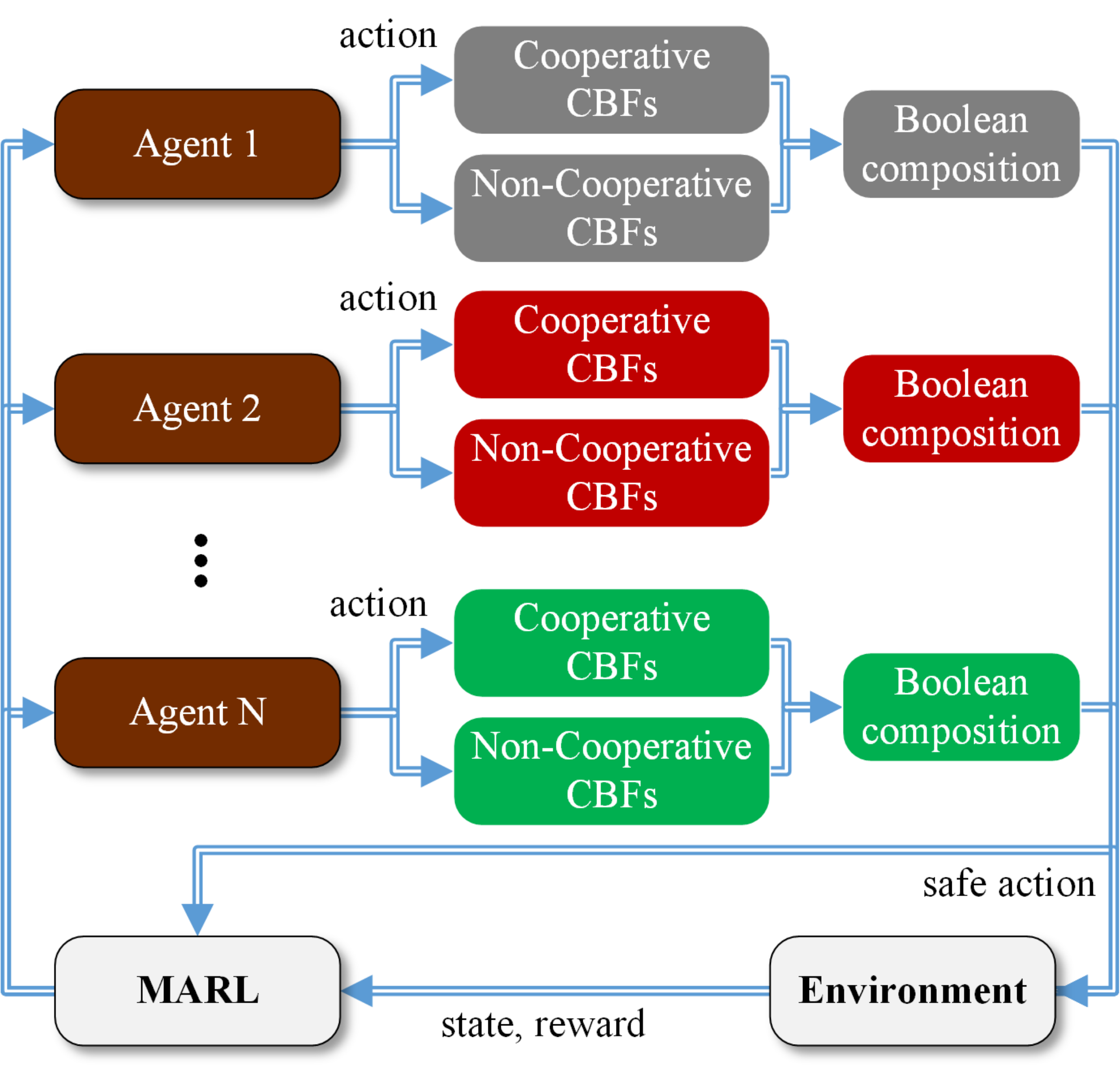}
     \caption{Safe MARL with decentralized multiple control barrier functions.}
     \label{flow_diagram}
\end{figure}

A centralized shield MARL has been proposed recently \cite{ElsayedAly2021SafeMR} to address the safety issues of a general MARL that does not have a specific mechanism for ensuring safety. Nonetheless, a centralized shield is infeasible for agents when suffering several scenarios such as non-cooperative agents and communication delay. Relative to the centralized shield, agents demand decentralized self-protection mechanisms designed for surviving from not only cooperative agents but non-cooperative agents more urgently. Thus, inspired by recent works \cite{Fisac2019ASystems,Cheng2019End-to-endTasks} that involve model information into RL to guarantee safety, we propose a safe MARL framework integrating decentralized multiple CBFs to agents of MARL, as shown in Fig. \ref{flow_diagram}. 

According to the proposed safe MARL framework, an agent has its own CBFs that can be different from the CBFs of other agents. Inspired by the fact that agents and entities in a multi-agent environment are not always cooperative, an agent in the proposed safe MARL framework is equipped with multiple CBFs, including cooperative CBFs and non-cooperative CBFs designed for cooperative agents and non-cooperative agents or entities, respectively. An agent that has $k$ CBFs can combine its CBFs through Boolean composition \cite{Glotfelter2017NonsmoothSystems} to achieve a composite CBF as
\begin{equation}
\label{Boolean}
\begin{split}
    B_c(x) &= B_1(x) \cap  B_2(x)\cap ... \cap  B_k(x) \\
\end{split}
\end{equation}
We expect that the composite CBF lead the minimal interference \cite{Fisac2019ASystems,ElsayedAly2021SafeMR}  to the action of an agent. Namely, 1) the composite CBF corrects the action of an agent only if it tends to violate safety constraints, and 2) the composite CBF revises the action of an agent as few as possible. A composite CBF derived based on (\ref{Boolean}) can lead to a safe action $u_{safe}$ for an agent as \cite{Ames2019ControlApplications}
\begin{equation}
\label{qp}
\begin{split}
&u_{safe}= \arg\min_u \frac{1}{2}\vert\vert u-\hat u\vert\vert^2\\
s.t. 
 \ \ \ & L_fB_c(x)+L_gB_c(x)u -\frac{\Gamma }{B_c(x)} \le 0 \\
\end{split}
\end{equation}
where $\hat{u}$ is the nominal action of the agent determined based on MARL. 

It should be emphasized that decentralized CBFs are designed for an agent based on its local information, so the dynamics of the environment and other agents do not have to be completely known beforehand. Moreover, decentralized CBFs can not only enforce the safety constraints during the learning process and the deployment process, but also improve the learning efficiency according to \cite{Cheng2019End-to-endTasks}.

 
Based on the proposed safe MARL framework, we develop MADDPG to MADDPG-CBF, a safe MARL algorithm. The MADDPG-CBF can be expressed as Algorithm 1. At time step $t$, agent $i$ selects an action $u^i_t$ based on available information $x^i_t$ and policy $\pi_{\theta_i}^{i}$. With the action $u^i_t$, agent $i$ can achieve a safe action $u^i_{t,safe}$ according to (\ref{Boolean}) and (\ref{qp}).


\begin{algorithm}[htb]  
	\caption{MADDPG-CBF Algorithm}  
	\label{alg:Framwork}  
	\begin{algorithmic}[0]
		\State \textbf{Initialize}
        \State{\qquad Initialize replay buffer $\mathcal{D}$}
		\State{\textbf{For episode = 1 to $m$ do}:}
		\State{\qquad Reset environment}
		\State{\qquad Initialize state $x$}
		\State{\qquad \textbf{For time step $t$ = 1 to max-episode-length do}}
		\State{\qquad \qquad Initialize a random process $\mathbb{N}_t$ for action \\
		\qquad \qquad  exploration}
		\State{\qquad \qquad \textbf{For agent $i$ = 1 to $n$ do}}
		\State{\qquad \qquad \qquad Select action $u^i_{t}=\pi_{\theta_i}^{i}(x_t^i) + \mathbb{N}_t$ w.r.t\\ \qquad \qquad \qquad the current policy and exploration}
		\State{\qquad \qquad\qquad Determine the composite CBF $B_c^i(x)$ \\ \qquad \qquad \qquad according to (\ref{Boolean})}
		
		\State{\qquad \qquad \qquad Obtain safety action $u^i_{t,safe}$ based on \\ \qquad \qquad \qquad the nominal action $u^i_{t}$ according to (\ref{qp})}
        \State{\qquad \qquad \qquad Update action $u^i_t \leftarrow u^i_{t,safe}$ }\\
        \qquad\qquad\textbf{End For}
		\State{\qquad \qquad  Execute actions ${\{u^i_t\}}_{i \in N}$} 
		\State{\qquad \qquad  Observe rewards ${\{r^i_t\}}_{i \in N}$ and new state $x'$}
		\State{\qquad \qquad Store $(x, {\{u^i_t\}}_{i \in N}, {\{r^i_t\}}_{i \in N}, x')$ in replay buffer $\mathcal{D}$}
		\State{\qquad \qquad  Update state $x \leftarrow x'$}
		\State{\qquad \qquad \textbf{For agent i =1 to N do}:}
		\State{\qquad \qquad\qquad Sample a random minibatch of $S$ samples \\
		\qquad\qquad \qquad $(x_j, u_j, r_j, x'_j)$ from $\mathcal{D}$} 
		\State{\qquad \qquad\qquad Calculate the target state-action value:\\
		\qquad \qquad
		$y_j=r^i_j+\gamma Q^i_{\pi^\prime}\left(x^{\prime }_j,u^{\prime 1},\ldots,u^{\prime N}\right)|_{\ u^{\prime i}=\pi^{\prime i}(x_j^i)}$}
		\State{\qquad \qquad \qquad Update critic by minimizing the loss:\\
		\qquad  \qquad$L\left(\theta_i\right)=\frac{1}{S}\sum_{j}{(y_j-Q^i_\pi\left(x_j,u^1_j,\ldots, u^N_j\right))}^2$}
		\State{\qquad \qquad \qquad Update actor using the gradient of loss:\\
	    \qquad  $\nabla_{\theta_i}J\approx\frac{1}{S}\sum_{j}{\nabla_{\theta_i}\pi^i(x^i_j)\nabla_{\theta_i}Q^i_{\pi}\left(x_j,  u^1_j,\ldots,u^N_j\right)}_{u^i=\pi^i(x^i_j)}$}
		\State \qquad \qquad \qquad Update target network parameters:\\ 
		\qquad \qquad \qquad \qquad  $\theta'_i\leftarrow\xi\theta_i+(1-\xi)\theta'_i$
		\State{\qquad\qquad \textbf{End For}}
		\State{\qquad \textbf{End For}}
		\State \textbf{End For}

	\end{algorithmic}  
\end{algorithm}  

\section{Decentralized Control Barrier Functions in Collision-Avoidance}
To further demonstrate the decentralized multiple CBFs included in the proposed safe MARL framework, cooperative CBFs and non-cooperative CBFs designed for a cooperative multi-robot system worked in an environment with obstacles are proposed in this section. A multi-robot system can involve a number of agents, leading to significant challenges in terms of collision-avoidance. The key to ensuring safety is that all potential pairwise agent-to-agent and agent-to-obstacle collisions are accounted for \cite{Wang2017SafetySystems}. Thus, pairwise agent-to-agent and agent-to-obstacle collisions that are nearby an agent are concerned by the agent based on available local information. CBFs are designed for an agent accounting for nearby pairwise agent-to-agent and agent-to-obstacle collisions.

It is assumed that available local information for an agent includes the position and velocity of nearby agents and obstacles as well as the acceleration of nearby agents. Since the multi-robot system is cooperative, cooperative CBFs are designed for the agent accounting for agent-to-agent collisions. Non-cooperative CBFs are designed for the agent accounting for agent-to-obstacle collisions due to the fact that obstacles will not cooperate with the agent to avoid collisions. Based on these multiple cooperative and non-cooperative CBFs, the agent can achieve a safe action according to (\ref{Boolean}) and (\ref{qp}).





\begin{figure}[h]

    \centering
    \includegraphics[width=4.5 cm]{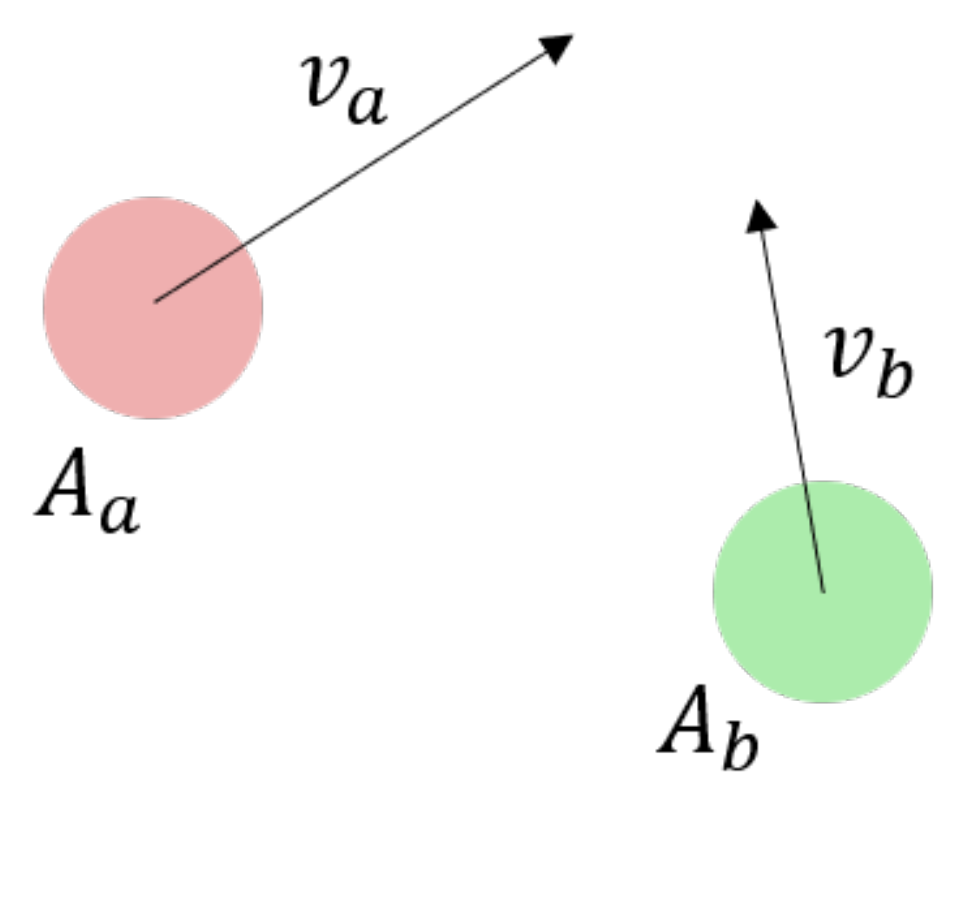}
    \caption{A pair of agent $A_a$ and agent $A_b$}
    \label{velocity}
 \end{figure}

Although both pairwise agent-to-agent and agent-to-obstacle collisions are considered, we regard obstacles as agents with certain non-cooperative behaviors for the convenience of expression. For a pair of two agents, denoted as $A_a$ and $A_b$ as shown in Fig. \ref{velocity}, a cooperative CBF or a non-cooperative CBF is designed for the agent $A_a$. The states of agents $A_a$ and $A_b$ are represented by $x_a = [p_a, v_a]^T$ and $x_b = [p_b, v_b]^T$. $p_a$ and $p_b$ are the position vectors of agents $A_a$ and $A_b$, respectively. $v_a$ and $v_b$ represent the velocity vectors of agents $A_a$ and $A_b$, respectively.
Assume that the dynamics of agents $A_a$ and $A_b$ is expressed as \cite{Wang2017SafetySystems}
\begin{equation}
\label{dynamics1}
    \dot{x}_a=
    \begin{bmatrix}
            \dot{p}_a\\
            \dot{v}_a\\
    \end{bmatrix} =
    \begin{bmatrix}
            0&I\\
            0&0\\
    \end{bmatrix}
    \begin{bmatrix}
            p_a\\
            v_a\\
    \end{bmatrix}
    +\begin{bmatrix}
            0\\
            I\\
    \end{bmatrix}
    a_a
\end{equation}

\begin{equation}
\label{dynamics2}
        \dot{x}_b=
    \begin{bmatrix}
            \dot{p}_b\\
            \dot{v}_b\\
    \end{bmatrix} =
    \begin{bmatrix}
            0&I\\
            0&0\\
    \end{bmatrix}
    \begin{bmatrix}
            p_b\\
            v_b\\
    \end{bmatrix}
    +\begin{bmatrix}
            0\\
            I\\
    \end{bmatrix}
    a_b
\end{equation}
where $a_a$ and $a_b$ are acceleration vectors of agents $A_a$ and $A_b$. $I$ is unit matrix.

\subsection{Cooperative Control Barrier Function}

In this subsection, a cooperative CBF is designed for agent $A_a$ according to \cite{Borrmann2015ControlBehavior}, based on the assumption that agents $A_a$ and $A_b$ will cooperate to avoid collisions. Let $\Delta p$ represent the vector of the position difference between agent $A_a$ and agent $A_b$. $\Delta v$ represents the vector of the velocity difference between agent $A_a$ and agent $A_b$. A cooperative safe set is defined as
\begin{equation}
\label{coo}
    C_{coo}=\left \{(\Delta p, \Delta v) \ | \ -\frac{\Delta p^T}{\lvert \lvert \Delta p \rvert \rvert} \Delta v \leq \sqrt{2 \Delta a_{max}(\lvert \lvert \Delta p \rvert \rvert-d_s)} \right \}
\end{equation}
where $d_s$ is a predefined safe distance. $\Delta a_{max} = a_{amax}+a_{bmax}$ is a constant defined by the maximum acceleration of $A_a$, denoted as $a_{amax}$, and the maximum acceleration of $A_b$, denoted as $a_{bmax}$. A continuously differentiable function $h_{coo}(x)$ is defined as
\begin{equation}
\label{hcoo}
    h_{coo}(x)=\frac{\Delta p^T}{\lvert \lvert \Delta p \rvert \rvert} \Delta v + \sqrt{2 \Delta a_{max}(\lvert \lvert \Delta p \rvert \rvert-d_s)}
\end{equation}
A candidate cooperative CBF is designed as
\begin{equation}
\label{bcoo}
    B_{coo}(x) = \frac{1}{h_{coo}(x)}
\end{equation}
According to (\ref{B}), linear constraints on cooperative actions of agents $A_a$ and $A_b$ can be expressed as
\begin{equation}
\label{constrains}
    \begin{split}
        -\Delta p^T \Delta a &\leq \frac{\Gamma_{coo}}{B_{coo}}h_{coo}^2\lvert \lvert \Delta p \rvert \rvert - \frac{(\Delta v^T \Delta p)^2}{\lvert \lvert \Delta p \rvert \rvert ^2}\\
        &+\lvert \lvert \Delta v \rvert \rvert ^2 + \frac{\Delta a_{max}\Delta v^T \Delta p}{\sqrt{2 \Delta a_{max}(\lvert \lvert \Delta p \rvert \rvert-d_s)}}
    \end{split}
\end{equation}
where $\Delta a = a_a-a_b$, $\Gamma_{coo} > 0$. Linear constraints in (\ref{constrains}) can be written as 
\begin{equation}
\label{Au1}
    A_{coo}u_{coo} \leq b_{coo}
\end{equation}
where $A_{coo} = -\Delta p^T$ and $u_{coo}=\Delta a$. $b_{coo}$ represents the right side of (\ref{constrains}).

According to \textbf{Theorem 1}, if $u_{coo}$  satisfies  (\ref{Au1}) as well as $\lvert\lvert u_{coo}\rvert\rvert \leq \Delta a_{max}$, then $B_{coo}(x)$ is a valid CBF and the safe set $C_{coo}$ is forward invariant, ensuring the safety of agents $A_a$ and $A_b$.  

\subsection{Non-Cooperative Control Barrier Function}

In this subsection, we develop a non-cooperative CBF for agent $A_a$ accounting for nearby agent-to-obstacle collisions. Since the non-cooperative CBF accounts for agent-to-obstacle collisions, we have $v_b=0$, $a_b=0$, and $a_{bmax}=0$. $p_b$ is a constant. We define a non-cooperative safe set as
\begin{equation}
      C_{noo}=\left \{(\Delta p, v_a) \ | \ -\frac{\Delta p^T}{\lvert \lvert \Delta p \rvert \rvert} v_a \leq \sqrt{2 a_{amax}(\lvert \lvert \Delta p \rvert \rvert-d_s)} \right \}
\end{equation}
A continuously differentialable function $h_{non}(x)$ is defined as
\begin{equation}
      h_{coo}(x)=\frac{\Delta p^T}{\lvert \lvert \Delta p \rvert \rvert} v_a + \sqrt{2 a_{amax}(\lvert \lvert \Delta p \rvert \rvert-d_s)}
\end{equation}
Based on $h_{non}(x)$, we construct a candidate non-cooperative CBF as

\begin{equation}
\label{bnon}
    B_{non}(x) = \frac{1}{h_{noo}(x)}
\end{equation}
According to (\ref{B}), linear constraints on action of agent $A_a$ can be expressed as

\begin{equation}
\label{non_constraint}
    \begin{split}
        -\Delta p^T a_a &\leq \frac{\Gamma_{non}}{B_{non}}h_{non}^2\lvert \lvert \Delta p \rvert \rvert - \frac{(v_a^T \Delta p)^2}{\lvert \lvert \Delta p \rvert \rvert ^2}\\
        &+\lvert \lvert v_a \rvert \rvert ^2 + \frac{ a_{amax} v_a^T \Delta p}{\sqrt{2 a_{amax}(\lvert \lvert \Delta p \rvert \rvert-d_s)}}
    \end{split}
\end{equation}
where $\Gamma_{non} >0$. Linear constraints in (\ref{non_constraint}) also can be expressed as
\begin{equation}
\label{Au2}
    A_{non}u_{non} \leq b_{non}
\end{equation}
where $A_{non} = -\Delta p^T$ and $u_{non}=a_a$. $b_{non}$ represents the right side of (\ref{non_constraint}).

According to \textbf{Theorem 1}, if $u_{non}$  satisfies (\ref{Au2}) and $\lvert\lvert u_{non}\rvert\rvert \leq a_{amax}$, then $B_{non}(x)$ is a valid CBF and the safe set $C_{non}$ is forward invariant, ensuring the safety of agent $A_a$.  



\section{Experiments}

\begin{figure}[h]
     \centering
     \includegraphics[width=8.5 cm]{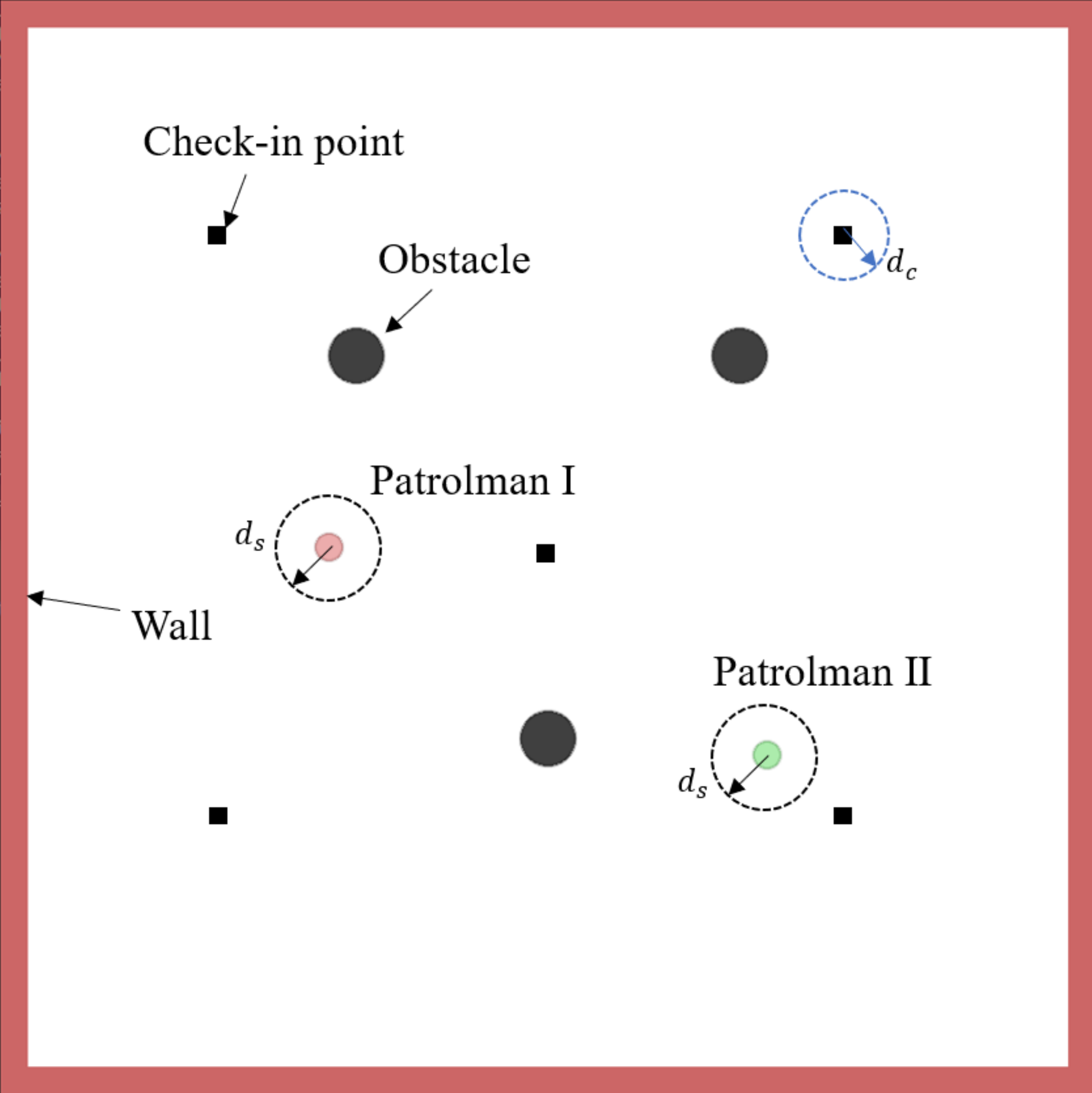}
     \caption{An environment consists of agents, obstacles, and walls.}
     \label{Environment}
 \end{figure}

In this section, we implement the proposed safe MARL framework with decentralized multiple CBFs in the Multi-agent Particle Environment \cite{Lowe2017Multi-agentEnvironments}. We apply our demo implementation to patrol tasks conducted by two agents in an environment with obstacles, as shown in Fig. \ref{Environment}. 

\textbf{Problem setup.} According to Fig. \ref{Environment}, the environment includes two agents, three static obstacles, and a square wall. The length of the side of the square wall is 2. The two agents are supposed to conduct different patrol tasks. One should perform random patrol tasks, denoted as Patrolman $\RNum{1}$, and the other should perform the task of checking in and patrolling, denoted as Patrolman $\RNum{2}$. Patrolman $\RNum{2}$ is expected to reach a predefined critical distance, denoted as $d_c$, around five check-in points represented by black rectangles in Fig. \ref{Environment} in order. $d_c$ is set to 0.05 in the experiment. The dynamics of Patrolman $\RNum{1}$ and Patrolman $\RNum{2}$ accords with (\ref{dynamics1}) and (\ref{dynamics2}). The maximum acceleration of agents is 1. The safe distance $d_s$ from an agent to the other agent or an obstacle is set to 0.075. The initial positions of Patrolman $\RNum{1}$ and Patrolman $\RNum{2}$ are randomly initialized. 


In the experiment, we use MADDPG as a baseline and implement MADDPG-CBF to show the effectiveness of the purposed safe MARL framework. In the implementation of MADDPG-CBF, actions of both Patrolman $\RNum{1}$ and Patrolman $\RNum{2}$ are monitored by their own cooperative CBFs and non-cooperative CBFs achieved according to (\ref{bcoo}) and (\ref{bnon}), respectively. Both algorithms are used to train policies based on reward functions defined as follows. 

The reward function of Patrolman $\RNum{1}$ is defined as


\begin{equation}
    r_{\RNum{1}}= \sum_{l=1}^{k}\begin{cases}
        -50& \text{$d_{\RNum{1}}^l \le d_s$}\\
        50& \text{$d_{\RNum{1}}^l > d_s$}\\
    \end{cases}
    \label{reward1}
\end{equation}
where $d_s$ is the safe distance and $d_{\RNum{1}}^l$ represents the distance between Patrolman $\RNum{1}$ and the $l$th nearby entity. An entity can be the other agent or an obstacle. $k$ denotes the number of nearby entities. 


The reward function of Patrolman $\RNum{2}$ is defined as


\begin{equation}
    r_{\RNum{2}}= \sum_{l=1}^{k}\begin{cases}
        -50 & \text{$d_{\RNum{2}}^l \le d_s$}\\
        50 & \text{$d_{\RNum{2}}^l > d_s$  and $d_{\RNum{2}}^c > d_c$} \\
        100 & \text{$d_{\RNum{2}}^l > d_s$  and $d_{\RNum{2}}^c \leq d_c$} \\
    \end{cases}
    \label{reward2}
\end{equation}
where $d_{\RNum{2}}^l$ is the distance between Patrolman $\RNum{2}$ and the $l$th entity. $d_{\RNum{2}}^c$ represents the distance between Patrolman $\RNum{2}$ and a given target check-in point. 

\begin{figure}[h]
     \centering
     \includegraphics[width=8.6cm]{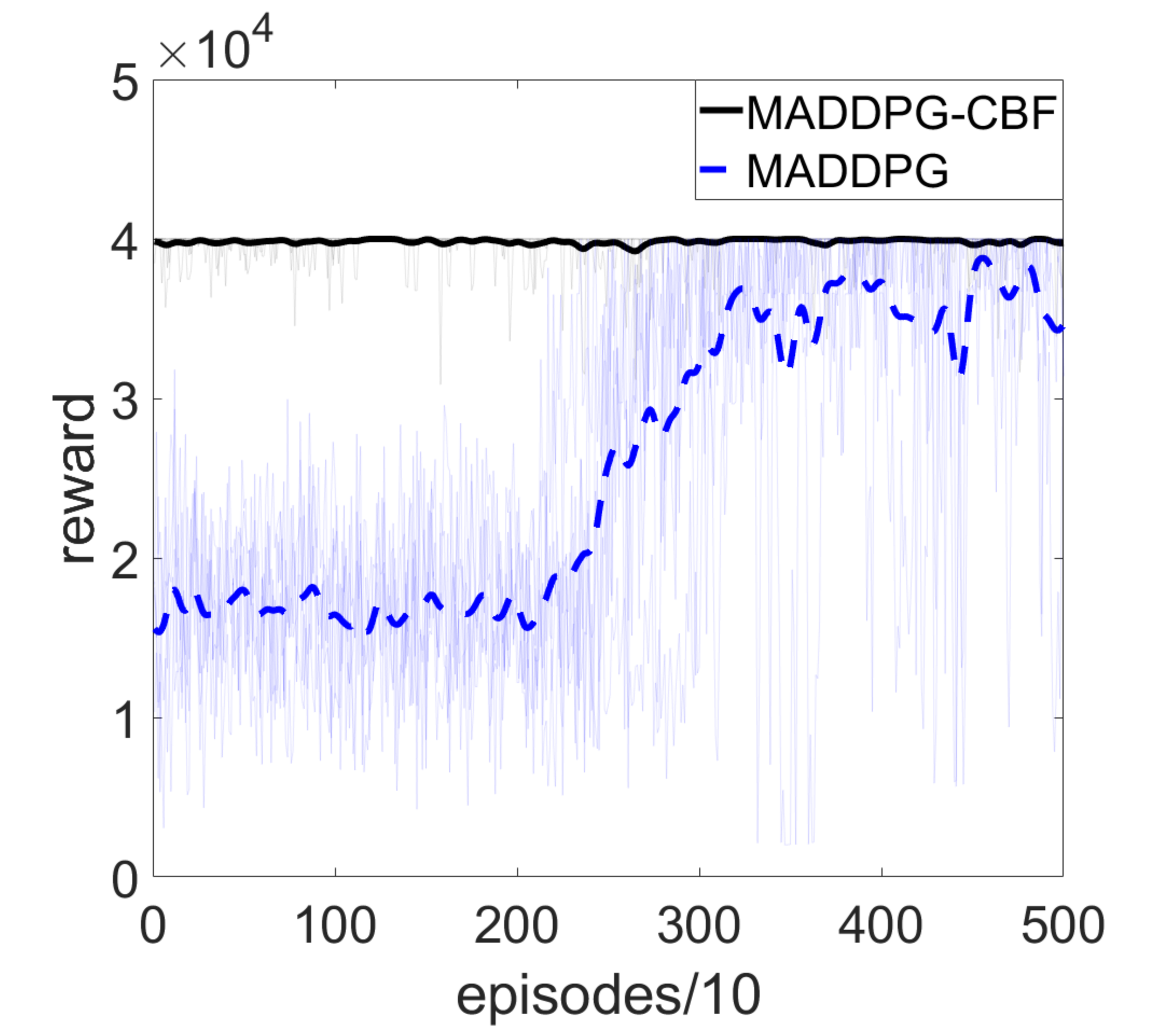}
     \caption{Average total rewards of two agents achieved by MADDPG and MADDPG-CBF.}
     \label{reward}
 \end{figure}

The average total rewards of two agents achieved by these two algorithms in five training runs are shown in Fig. \ref{reward}. It is shown that the average total reward achieved by MADDPG-CBF stabilizes at around 40,000 from the beginning of the training, while the average total reward achieved by MADDPG converges in 3,200 episodes. To guide the training of policies that tend to avoid collisions, the reward functions of agents expressed in (\ref{reward1}) and (\ref{reward2}) heavily depend on the distance from an agent to its nearby entities. Thus, with CBFs that are designed to address collisions, the average total reward of MADDPG-CBF can stabilize at a large reward from the beginning of the training.


The occurrences of collisions in the five 5,000-episode training runs are studied and expressed in Table I. The results in Table I show that MADDPG-CBF can guarantee collision-free learning in all episodes. On the contrary, MADDPG without decentralized multiple CBFs suffers 12,634 occurrences of collisions, suggesting that, without decentralized multiple CBFs, MADDPG leads to collisions. 

\begin{table}[h]
\centering
    \begin{tabular}{|c|c|c|}
    \hline
    \rule[-1pt]{0pt}{10pt}   & MADDPG-CBF & MADDPG \rule[-1pt]{0pt}{10pt}  \\ \hline
    \rule[-1pt]{0pt}{15pt} Number of collisions  & 0  &   12,634 \rule[-1pt]{0pt}{15pt}     \\ \hline
    \rule[-1pt]{0pt}{15pt} Number of episodes  & 25,000  &   25,000 \rule[-1pt]{0pt}{15pt}     \\ \hline
    \rule[-1pt]{0pt}{15pt} Collision ratio &0$\%$  &   50.536$\%$  \rule[-1pt]{0pt}{15pt}    \\ \hline
    \end{tabular}
    \caption{ Number of collisions and collision ratio in five training runs.}
\end{table}

\begin{figure}[h]
     \centering
     \includegraphics[width=7.5cm]{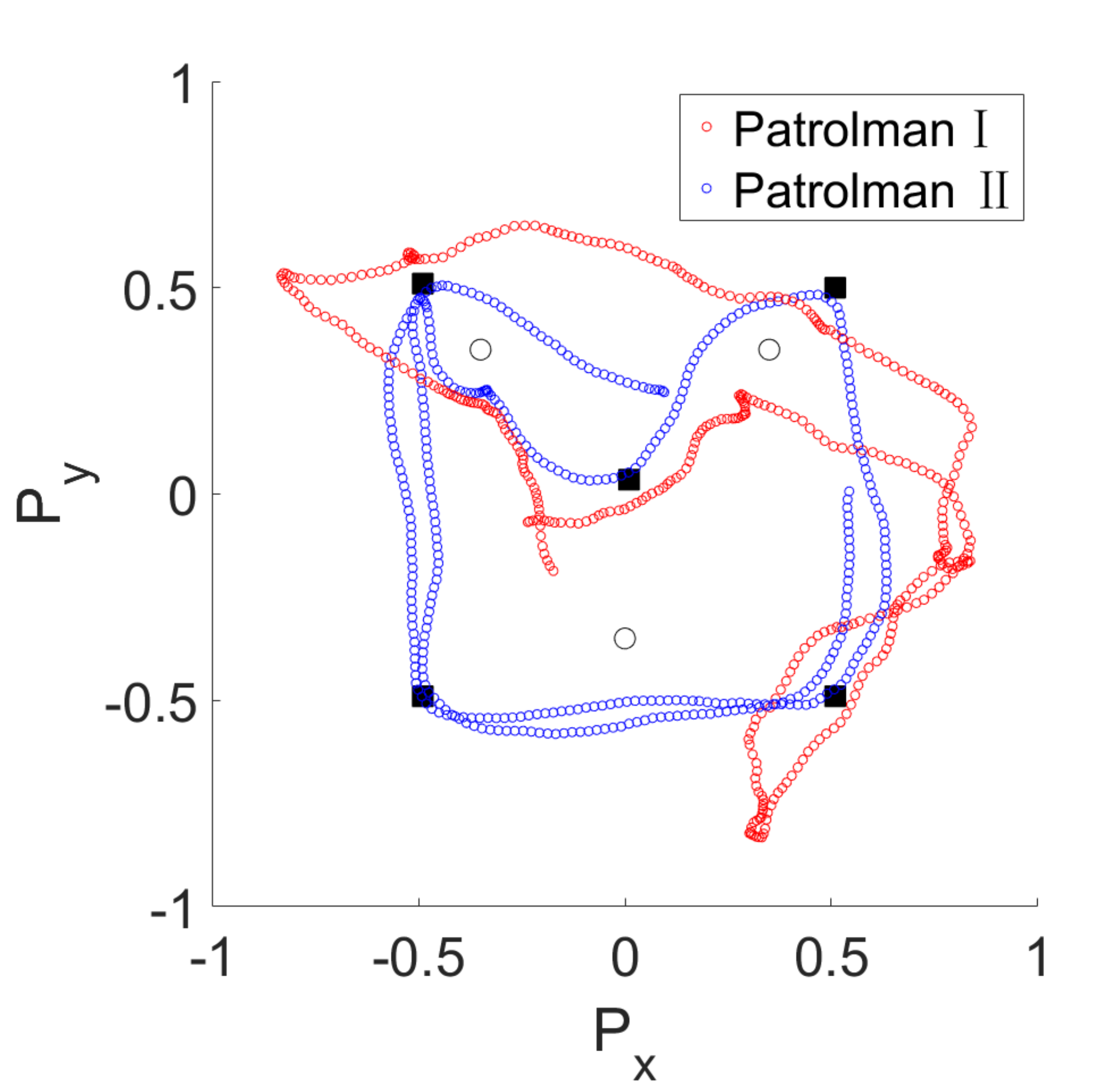}
     \caption{Trajectories of two patrolmen in one episode.}
     \label{trajectries}
\end{figure}

We analyze one of the last episodes of training based on MADDPG-CBF in detail. Trajectories of agents in the episode are shown in Fig. \ref{trajectries}. One can see that Patrolman $\RNum{1}$ and Patrolman $\RNum{2}$ are possible to suffer collisions in this environment if CBFs are not available. Moreover, Patrolman $\RNum{2}$ reaches all five check-in points in the episode as expected. Patrolmen can accomplish their patrol tasks while avoiding collisions. 


\begin{figure}[h]
     \centering
     \includegraphics[width=7.5cm]{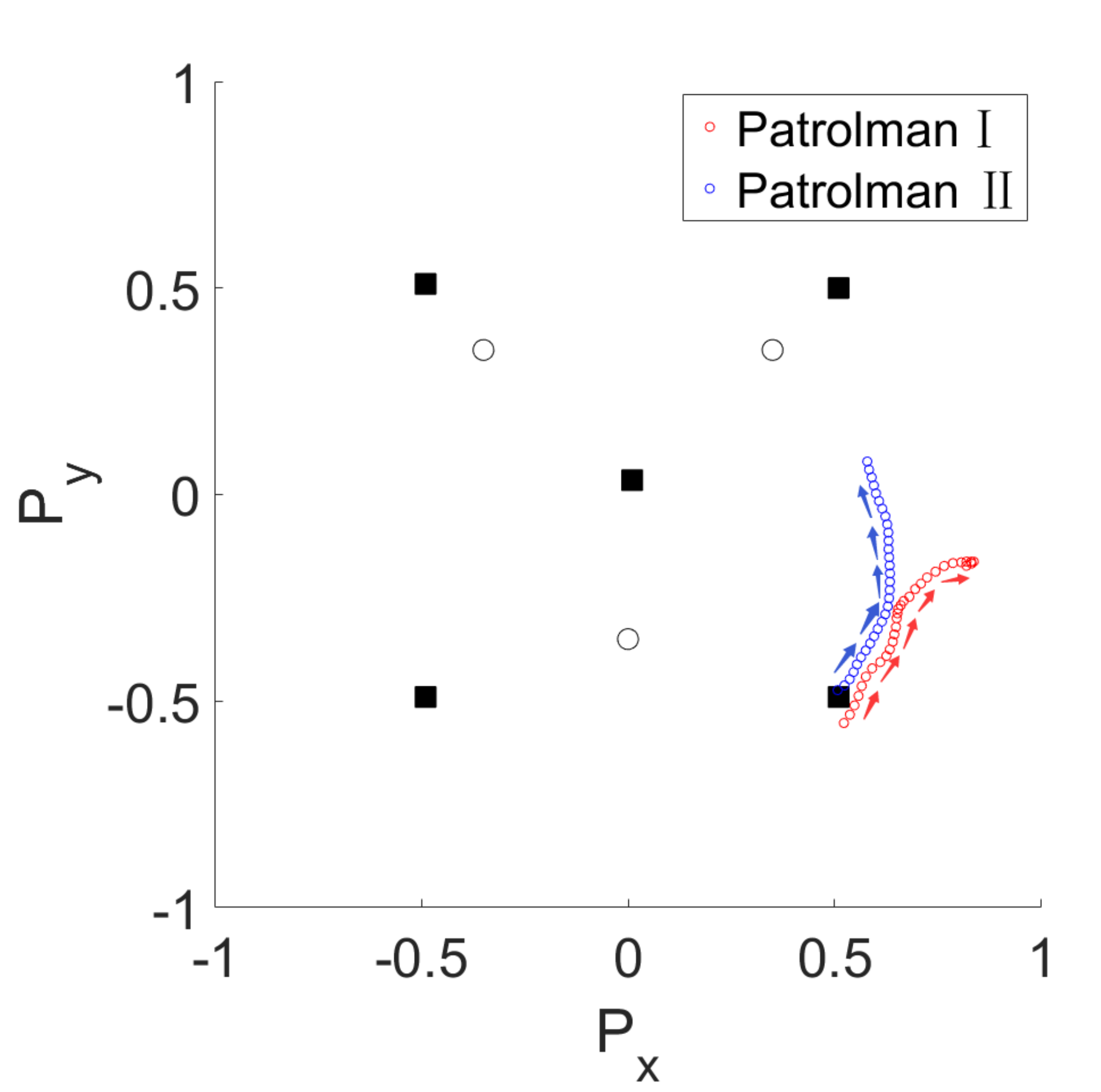}
     \caption{Trajectories of two patrolmen in the first segment of the episode.}
     \label{segement1}
\end{figure}

\begin{figure}[h]
     \centering
     \includegraphics[width=7.5cm]{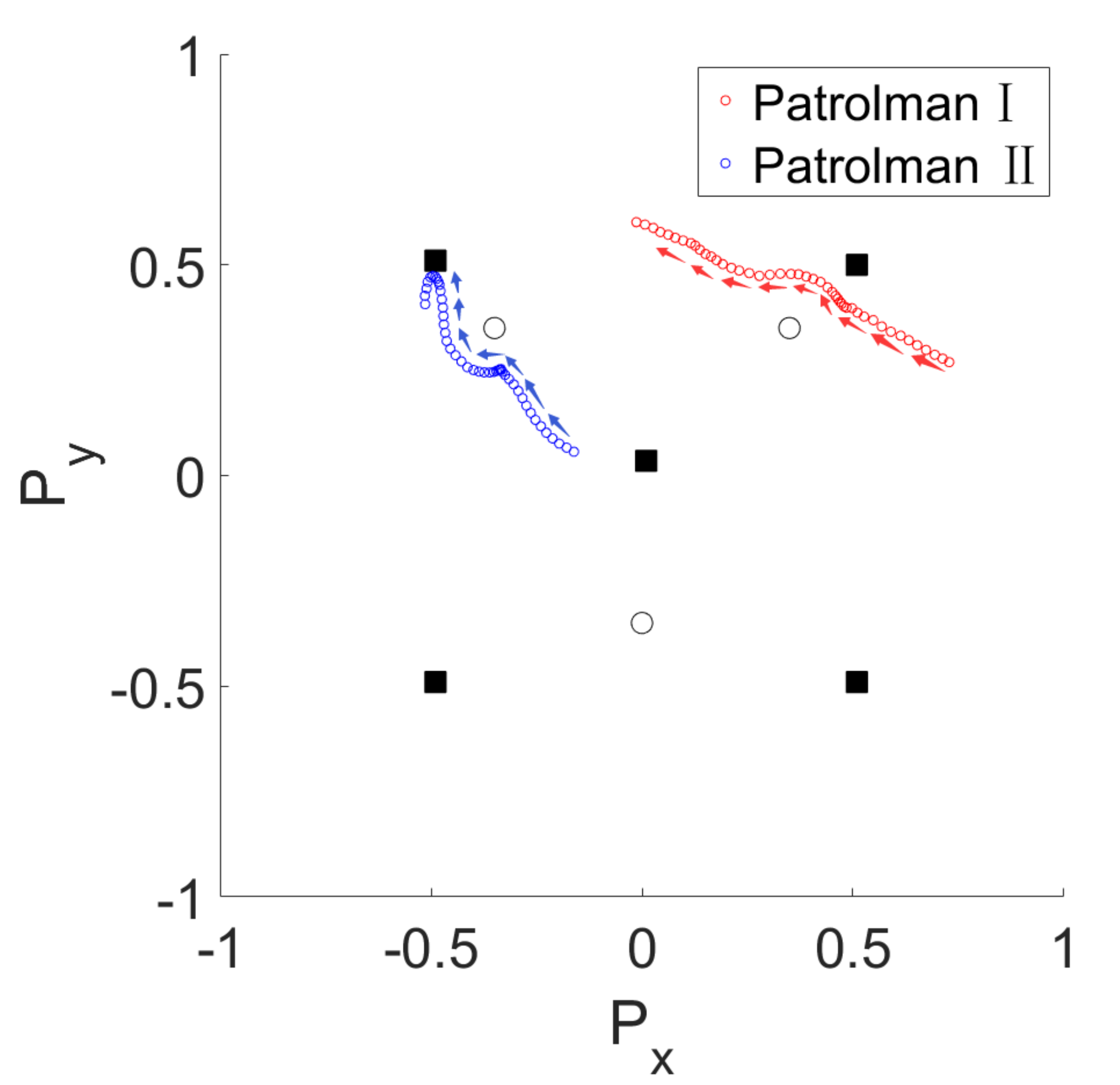}
     \caption{Trajectories of two patrolmen in the second segment of the episode.}
     \label{segement2}
\end{figure}

We study the behaviors of agents in the episode shown in Fig. \ref{trajectries} as well. Trajectories of agents in two segments of the episode are shown in Figs. \ref{segement1} and \ref{segement2}. Fig. \ref{segement1} shows that, when Patrolman $\RNum{1}$ and Patrolman $\RNum{2}$ move close to each other, they can steer their headings to avoid an agent-to-agent collision. According to Fig. \ref{segement2}, Patrolman $\RNum{2}$ can go round an obstacle when approaching a check-in point and Patrolman $\RNum{1}$ can get round as well when heading to an obstacle initially. The behaviors of agents in these segments verify the safety of MADDPG-CBF.



\section{Conclusions and Future Work}

MARL bears safety concerns when used in safety–critical applications. In this paper, we proposed a safe MARL framework, in which decentralized multiple CBFs are integrated to every single agent. According to the safe MARL framework, MADDPG-CBF was developed based on MADDPG. Through an introduction of cooperative and non-cooperative CBFs in a collision-avoid problem, this paper demonstrated that the safe MARL framework is with safety guarantees in certain applications. Experiments of MARL in an environment that includes obstacles were conducted to validate the effectiveness of the safe MARL framework. Experiment results show that, based on MADDPG-CBF, agents can avoid agent-to-agent and agent-to-obstacle collisions, while agents suffer collisions based on MADDPG. In the future, we will focus on the balance between the safety and the performance of MARL in a multi-agent multi-objective environment.





\input{root.bbl}

\bibliographystyle{IEEEtran}

\end{document}